# SIMULATED ANNEALING FOR LOCATION AREA PLANNING IN CELLULAR NETWORKS

N. B. Prajapati[1], R. R. Agravat[2] and M. I. Hasan[3]

[1]I. T. Department, B. V. M. Engineering. College, V. V. Nagar- 388120, India
*nbp_it53@yahoo.com*
[2]I. T. Department, B. V. M. Engineering. College, V. V. Nagar- 388120, India
*rupalagravat@yahoo.com*
[3]C. P. Department, B. V. M. Engineering. College, V. V. Nagar- 388120, India
*mosin83@yahoo.com*

*ABSTRACT*

*LA planning in cellular network is useful for minimizing location management cost in GSM network. In fact, size of LA can be optimized to create a balance between the LA update rate and expected paging rate within LA. To get optimal result for LA planning in cellular network simulated annealing algorithm is used. Simulated annealing give optimal results in acceptable run-time.*

*KEYWORDS*

*Location Area, Location Update, Base Station, Mobile Station, Base Station Controller, Mobile Switching Center, Simulated Annealing*

## 1. INTRODUCTION

Everyone wants mobility and because of that the wireless concept is growing up and up and wireless network is changing day by day. The First and Second generation wireless networks, such as analog cellular, Global System for Mobile Communication and Personal Communication Systems have been widely deployed in the last decade or so.

Cellular networks are a type of networks that support wireless connection to mobile station (MS). Cellular networks is rapidly growing network in today's world because of a rapid growth in the population of mobile users, and such growth increases signalling. Signalling requires too much bandwidth for all operation, while Radio bandwidth is limited resource. To decrease signalling operations it is required to do better Mobile Network planning.

## 2. LOCATION AREA

In second generation cellular network, like GSM, the system coverage area is divided into geographical area called LA. This geographical area contains one or more cell areas called Base Station (BS). Each of this LA is identified by its unique Location Area Identifier (LAI). This LAI is broadcasted by each BS which includes all BSs forming that particular LA. The LAI consist three parts: a Mobile Country Code, a Mobile Network Code, and a 2-octect Location area code. *Figure II.1* shows the service area which is divided into three location areas, where dark lines show boundary of each LA while dotted lines show boundary of BS which are within that LA. In GSM network, LA and Paging area is considered similar.

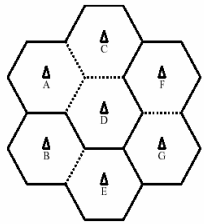

Figure.2-1: Service area divided into three Location Areas. [9]

There are two types of LA, described below:

| Type of LA | Advantage | Disadvantage |
|---|---|---|
| *Static* | Reduced Overhead | Same for all users with different call to mobility ratio |
| *Dynamic* | Based on user mobility patterns | Difficult to determine the optimal shape and size |

## 2.1. Importance of LA

There are different location update schemes which are used for location management in mobile networks. All cellular networks use the zone based approach for location management. In zone base approach the entire cellular network is divided in to LA [7]. So, LA is important factor in cellular network for Location management. Location management involves two basic operations: *Location Update* (LU) and *Paging*. In the Location update, also called Location Registration, the MS from time to time, notifies the network of its new access point, and the network database stores the new position, or LA. Paging is performed base on the location of MS in particular LA.

## 2.2. Size and Shape of LA

Size and Shape of LA is more important in Cellular network. The size of an LA or the number of cells in it may vary depending on the rate at which cells receive calls, and on the inter-cell traffic characteristics and number of user. In fact, the size of an LA can be optimized to create a balance between the LA update rate and the expected paging rate within an LA. In other words proper planning of LA reduces total Location management cost. There ate two extreme LA planning approaches:
1) The service area equals an LA: Here LUs are not required which generated due to mobile user movement in service area while paging cost is increased because Whenever an MS is called, it is paged over the whole service area. In this case paging signalling load can be enormous especially during the busy hour.
2) The cell area equals an LA: In this case the location of an MS is determined with accuracy of a single cell area. The need for paging here is minimal; paging does not locate the MS, it just alerts the terminal for the incoming call. However, the number of LUs is expected to be enormous due to the small cell size and the user mobility.

From these two approaches we come to known that upper bound on size of LA is the service area of an MSC and upper bound on number of LAs is numbers of BSs.

## 3. SIMULATED ANNEALING

Simulated annealing (SA) is a random-search technique which exploits an analogy between the way in which a metal cools and freezes into a minimum energy crystalline structure (the annealing process) and the search for a minimum in a more general system; it forms the basis of an optimization technique for combinatorial and other problems [13]. In 1983 SA was developed for dealing with highly nonlinear problems. An important characteristic of the SA algorithm is that it does not require specialist knowledge about how to solve a particular problem. This makes the algorithm generic in the sense that it can be used in a variety of optimization problems without change the basic structure of the computation [18]. There are two important criteria for SA: Selection of Neighbor and Cooling schemes. Accuracy and optimization of SA depends on these two criteria.

In Neighboring Solution, collection of all the optimal and non-optimal solutions is called Solution Space. During annealing process, algorithm randomly selects any solution from these Solution space.

## 4. FORMULATION OF PROBLEM

We know that, Location management cost is depended on the LA. So LA planning plays important role in cellular network. LA planning can be done by SA. In this paper, GSM network is used as cellular network model. If we want LA planning in GSM network it is necessary to satisfy all constrains required for reducing the total location management cost [1] and from that we are able to get better GSM network.

### 4.1. Simulated Annealing for LA Planning

For LA planning, SA takes these constraints in to consideration [1], load and capacity information as an input and finds an optimal or near-optimal solution as a network topology which includes the assignment of BSs to LAs.

#### 4.1.1. Complexity of LA Planning Problem (Solution Space)

The solution space consists of both feasible and unfeasible solutions. Here, solution space consists of all possible network topologies, the BS-to-BSC, BSC-to-MSC and BS-to-LA assignments determine the complexity of the solution.

#### 4.1.2. Neighbor Solution

A feasible solution is a topology where all network nodes (BSs, BSCs, MSCs) are connected, and the LA borders are specified such that the constraints are satisfied. Therefore, a (feasible) neighbor solution may be generated using following move:

*Changing a BS-to-BSC assignment*: A BS to be moved is randomly chosen, and then a BSC is randomly selected among all BSCs in the proximity of that BS. Before executing this move, the capacity constraints affected by this BS-BSC connection are checked. If these constraints are not violated, the new BSC is assigned to the BS. Then, a random feasible LA is searched to assign that BS among the existing LAs residing in the new BSC. Because of the limitation that an LA cannot spread over multiple MSCs, instead of checking whether the new and the old BSC are connected to the same MSC, directly try to find an LA residing in the new BS. If no feasible LA is found, then a new LA is created for the new BSC. Next, all load updates resulting from that move are calculated on the network [1].

**4.1.3. Cooling Schedules**

A Cooling schedule consists of three important components: setting the starting (initial) temperature, the decision of when and how to decrease the temperature, and the decision of when to stop algorithm.

*1. Starting Temperature*

The starting temperature of the initial cost and the starting probability is 0.5 with the use of the equation $T = -\max \Delta E / \ln P_0$. Initial cost is calculated using below equation and with which the aim of LA planning is satisfied.

$$\text{Minimize} \sum_{i} \sum_{j} d_{ij} h_{ij}$$

Initial temperature calculation is very important, because if it is too high, then it may take too long to reach the result. And if the initial temperature is too small, then the algorithm does not have the freedom of making sufficient number of random moves to visit different local minima and stay in relatively deep one.

*2. Frozen Value*

It is said that if we want to get good result then we should reduce T in each iteration by 0.95. So with the use of the initial T and the $\Delta T$, we get the frozen Value by the equation:

$$\text{Frozen Value} = T / \Delta T \quad \text{where } \Delta T = 0.95$$

Amount of reducing temperature may differ according to the selection of the annealing schedule. Other criteria is when to decrease temperature for that, at specific temperature, if the total number of neighbors tested (not just the accepted ones) equals approximately to the neighborhood size (equilibrium), then the temperature is decreased.

*3. Ending Condition*

The algorithm must stop if it can not find better solution anymore. In LA planning, the SA algorithm stops when particular temperature/cost is satisfied numbers of times. During SA search, "the best solution recorded and presented when the algorithm stop".

## 5. IMPLEMENTATION

The entire variables related to GSM network are used in implementation of LA planning using SA. To make this experiment dynamic and reusable all the variables are taken in text file format. Here SA is used to satisfy GSM constraints for obtain better cellular network.

### 5.1. Stepwise Details

*1. Make an initial feasible topology.*

To start the algorithm initial feasible topology is selected. This topology satisfies all the constraints.

*2. Calculate solution cost for initial topology and setting temperature.*

Cost is important factor in SA algorithm. For calculating cost for this solution *handoffmatrix* and *InverseLABStopologymatrix* are used. After that this cost is used for setting initial temperature of algorithm according to the function:

initialTemp = -((1.0)*cost/Math.log(P0));

*3. Decreasing temperature according to the compilation of iteration.*

To find out optimal or near-optimal solution for the mentioned algorithm, it is necessary to run experiment until its temperature goes down to zero. In each iteration algorithm checks for better solution in solution space and it decreases temperature according to that.

4. *Generate new topology.*

New topology is based on neighborhood structure. In experiment BS-to-BSC assignment neighborhood structure is used for getting new topology for algorithm. In this assignment randomly BS and BSC is selected from *BSBSctopologymatrix* and creating new topology in that assignment of BSC and BS is made. For that all the parameter based on *BSBSctopologymatrix* which are affected by this assignment are changed. First of all the *BSBSctopologymatrix* is changed according to assignment of random BS to random BSC so other parameter like BHCA, call traffic and TRX are required to adjust according to new topology.

5. *Constraints checked for new topology.*

Before considering above change all the constraints related to that are checked. These constraints are related to *BSBSctopology, BScMSctopology* and call traffic constraints, like call traffic capacity and BHCA capacity.

6. *To assign new LA, paging constraints are checked.*

When constraints related to new topology of *BSBSctopologymatrix* are satisfied then a random feasible LA is searched to assign to that BS among the existing LAs residing in new BSC. After selecting random LA, it is required to change the *LABStoplogymatrix*. After that pagingCapacity constraint of BS and BSc is calculated using new PagingVecor_LA.

7. *New LA is created.*

When Constraints are violated then it is necessary to create new LA. Because of the limitation that an LA cannot spread over multiple MSCs, instead of checking whether the new and old BSC are connected to the same MSC, directly find a new LA residing in the new LA.

8. *Select new topology.*

After all the constraints are satisfied by new topology and new LA is created for that topology we calculated the cost of that topology. If cost is improved then this solution is considered as initial solution. Otherwise worse solution is selected. The probability of accepting a worse solution is:

$$P = e^{-\Delta E/t}$$

9. *Temperature decrease.*

After selecting new topology temperature is decreased. Perform the step 4 to 8 until optimal or near optimal solution is not found. Stop cooling when solution found or solution meets the stopping criteria.

10. *Save and display best solution.*

In each iteration, best solution is stored and when the cooling is stopped at that time display the best solution with minimum cost.

## 6. CONCLUSION

SA is one of the best heuristic methods for finding optimal or near-optimal solution in cellular network in tolerable run-times. SA gives the better LA planning in cellular network so total location management cost is reduced. The main aim of the experiment of better LA planning is fulfilled.

## 6.1. Future Work

Our future work includes development of the heuristic method which gives better LA planning when LAs are overlapping with each other in cellular network. We also want to work for the dynamic LA: individual's mobile user LA according to their mobility in cellular network, which is one of the interesting and unresolved research problems.